\documentclass[aps,reprint,article,prx]{revtex4-1}
\usepackage{amsmath}
\usepackage{dcolumn}
\usepackage{graphicx}
\usepackage{float}
\usepackage{color}
\usepackage{ulem}

\begin{document}

\title{Planckian Dissipation and non-Ginzburg-Landau Type Upper Critical Field in Bi2201}
\author{Qihao Zang$^{1}$, Zhengyan Zhu$^{1}$, Zuyu Xu$^{2}$, Shichao Qi$^{3}$, Haoran Ji$^{3}$, Yiwen Li$^{1}$, Jian Wang$^{3}$, Huiqian Luo$^{4}$, Hua-Bing Wang$^{2,5}$ and Hai-Hu Wen$^{1,\dag}$}

\affiliation{$^{1}$National Laboratory of Solid State Microstructures and Department of Physics, Collaborative Innovation Center of Advanced Microstructures, Nanjing University, Nanjing 210093, China}
\affiliation{$^{2}$Research Institute of Superconductor Electronics, Nanjing University, 210093 Nanjing, China}
\affiliation{$^{3}$International Center for Quantum Materials, School of Physics, Peking University, Beijing 100871, China}
\affiliation{$^{4}$Institute of Physics, CAS, Beijing 100190, China}
\affiliation{$^{5}$Purple Mountain Laboratories, Nanjing 211111, China}

\begin{abstract}
  Resistivity and Hall effect measurements have been carried out on a micro-fabricated bridge of Bi2201 single crystal at low temperatures down to 0.4 K under high magnetic fields. When superconductivity is crashed by a high magnetic field, the recovered ``normal state'' resistivity still shows a linear temperature dependence in low temperature region. Combining with the effective mass and the charge carrier density, we get a linear scattering rate $1/\tau = \alpha k_{B} T/\hbar$ with $0.77<\alpha<1.16$, which gives a strong evidence of the Planckian dissipation. Furthermore, our results reveal a new type of temperature dependence of upper critical field, $H_{c2}(T)=H^*\sqrt{(1-t)/(t+0.154)}$, which is totally different from the expectation of the Ginzburg-Landau theory, and suggests uncondensed Cooper pairs above $H_{c2}(T)$ line.
\end{abstract}

\maketitle
\section{Introduction}\label{section1}

In some optimally doped or overdoped cuprate superconductors, the resistivity in normal state shows a linear temperature dependence in a wide temperature region. This was first observed in Bi$_{2+x}$Sr$_{2-x}$CuO$_{6+\delta}$ (Bi2201)~\cite{Bi2201}. In this pioneer work, the $T$-linear dependence of in-plane resistivity shows up at temperatures from just above $T_\mathrm{c} \approx$ 7 K up to 700 K. This behavior strongly violates the picture of electron-phonon scattering in normal metals and thus has attracted enormous attention. This feature was later also observed in many other high-$T_c$ cuprate superconductors, such as Pr$_{2-x}$Ce$_{x}$CuO$_{4\pm\delta}$ (PCCO)~\cite{PCCO1,PCCO2,PCCO3}, La$_{2-x}$Ce$_{x}$CuO$_{4}$ (LCCO)~\cite{LCCO,JinK}, La$_{2-x}$Sr$_{x}$CuO$_{4}$ (LSCO)~\cite{LSCO}, La$_{1.6-x}$Nd$_{0.4}$Sr$_{x}$CuO$_{4}$ (Nd-LSCO)~\cite{Nd-LSCO1,Nd-LSCO2} and Bi$_{2}$Sr$_{2}$CaCu$_{2}$O$_{8+\delta}$ (Bi2212)~\cite{Bi2212}. It occurs also in organic~\cite{BaFeCoAs}, pnictide~\cite{BaFeCoAs,BaFe2(As1-xPx)2} superconductors, as well as in ruthenate~\cite{Sr3Ru2O7} and many heavy fermion compounds~\cite{MQPT,QCP,YbRh2Si2,QC}.

This $T$-linear resistivity is regarded as anomalous both in low and high temperature regions. In conventional metals, the resistivity usually shows a saturation in the high temperature region due to the frequent scattering by phonons in the Mott-Ioffe-Regel limit. However, in some materials this saturation of resistivity is absent, such as in Bi2201~\cite{Bi2201}. In the low temperature region, the electron-phonon scattering should give rise to a power law relation of resistivity, namely $\rho(T)\propto T^5$. This will change to a quadratic temperature dependence $\rho\propto T^2$ when a weak correlation is considered (The Landau-Fermi liquid expectation). Thus a $T$-linear dependence of resistivity in the low temperature region is a direct indication of an unconventional metallic state. In order to explain this phenomenon, several theoretical hypotheses have been proposed, such as the marginal Fermi liquid theory in low dimensions~\cite{Varma}, exotic fluctuations due to the quantum criticality~\cite{MQPT}, and so on. In some heavy fermion materials, the $T$-linear resistivity can be seen when they are tuned to the quantum critical point (QCP) by some external parameters~\cite{QCP}. Therefore, the $T$-linear resistivity is often associated with the scattering near the QCP. In the electron-doped cuprate superconductors, the $T$-linear resistivity was seen just above the QCP~\cite{PCCO2} where the long-range antiferromagnetic (AF) order vanishes~\cite{NCCO}, thus it may be induced by the extremely enhanced spin fluctuations. The same situation occurs in iron-based superconductors. However, in the hole-doped cuprate samples, the doping values where the $T$-linear resistivity occurs are generally far away from the point where the AF order ends~\cite{LSCO,Nd-LSCO3}. Instead, these doping levels are close to the critical doping level $p^*$ where the pseudogap phase ends~\cite{LSCO,Nd-LSCO3,Kawasaki}. Near this doping point, it was found that the Fermi surface measured by angle resolved photoemission spectroscopy (ARPES) changes the topology from hole-like to electron-like~\cite{ARPES}. Thus this can also be understood as an effect near the QCP point, here it may involve a quantum phase transition between a pseudogap based ground state to a full Fermi surface based one.

Despite enormous evidence accumulated for this $T$-linear resistivity in a variety of compounds, it remains however still as a puzzle. Recently, it was proposed that this may be understood as a novel dissipation governed by the many body effect in an entangled compressible quantum matter~\cite{Zaanen0,JanZaanen,Zaanen1}. This theory postulates that the dissipation in unconventional metals may be upper bounded by the universal scattering rate, namely the Planckian dissipation with the scattering rate $1/\tau = \alpha (k_B/\hbar)T$, where $\alpha$ is in the scale of unity, $k_B$ and $\hbar$ are the Boltzmann and reduced Planck constants, respectively. Some experiments have shown consistency with this prediction~\cite{Planckian,NatureTaillefer}. However, counterexamples were also found in electron-doped cuprates~\cite{LCCOGreene} in which the resistivity exhibits a $T$-linear dependence in low-$T$ region, but a quadratic temperature dependence in the intermediate temperature region, and the latter certainly violates the upper bound of Planckian dissipation. Thus this $T$-linear resistivity may be applicable only in the doping region close to QCP where exotic scattering is involved. To verify the interesting picture of Planckian dissipation, it is very crucial to precisely determine the scattering rate in a model system with superconductivity completely suppressed. In addition, it is also curious to know the nature of the ``normal state'' with this $T$-linear resistivity when superconductivity is killed by a high magnetic field.

In this paper, we report measurements of in-plane resistivity and Hall effect on a micro-fabricated bridge of the Bi$_{2+x}$Sr$_{2-x}$CuO$_{6+\delta}$ (x = 0.05, $T_c^{zero}\approx$ 6.7 K) single crystal. The advantage of using this system is that we can expose most of the phase diagram of magnetic field versus temperature. The well fabricated structure can also allow us to precisely determine the scattering rate in a wide temperature region. Our results support the postulation that the scattering rate reaches the Planckian limit~\cite{Planckian,JanZaanen,Zaanen1}, and the related ``upper critical field'' may be the point of turning the sample from superconducting state to the one with a mixture of uncondensed Cooper pairs and strongly renormalized quasiparticles, not the pair breaking field given by the Ginzburg-Landau theory.

\section{Experiment}\label{sec:2}

The Bi$_{2+x}$Sr$_{2-x}$CuO$_{6+\delta}$ single crystals were grown by the traveling-solvent floating-zone (TSFZ) method with an optical floating-zone furnace equipped with four ellipsoidal mirrors~\cite{LuoHQ}. To make a difference from the La-doped Bi$_2$Sr$_{2-x}$La$_x$CuO$_{6+\delta}$, our present sample is a Bi/Sr self-substituted one. The single crystals selected under a microscope were characterized by x-ray diffraction, magnetization and resistivity measurements. The magnetization measurements were operated using a SQUID-VSM-7T (Quantum Design). The selected single crystals were mechanically stripped and then made into a Hall bar by photolithography. The resistivity and Hall resistivity were measured by a physical property measurement system (PPMS-16T, Quantum Design). The magnetic field was applied perpendicular to the $ab$-plane of the Bi2201 sample during the measurements.

Since the samples are very brittle, cautions must be taken during the fabrication process. A Bi2201 crystal with an in-plane area of about $500\times500$ $\mu m^2$ and a thickness of about 5 $\mu m$ is firstly glued onto a sapphire substrate with epoxy. To protect the surface of the crystal, a 60-$nm$-thick gold film is deposited on the crystal immediately after cleaving. A microbridge with dimension of $50\times300$ $\mu m^2$ is patterned on a flat surface of the sample using photolithography and then the whole sample is etched down to the epoxy by argon ion milling. We remove the photoresist layer and subsequently deposit a 100-$nm$-thick gold film. Finally, the Hall bar electrodes are defined on the microbridge by photolithography and etched down to the crystal layer. After cleaving, the thickness of the sample is reduced from the beginning one.

\section{Experimental results}\label{sec:3}
\subsection{Resistivity and Hall effect}\label{sec:3.1}

\begin{figure}[H]
\centering
\includegraphics[width=8cm]{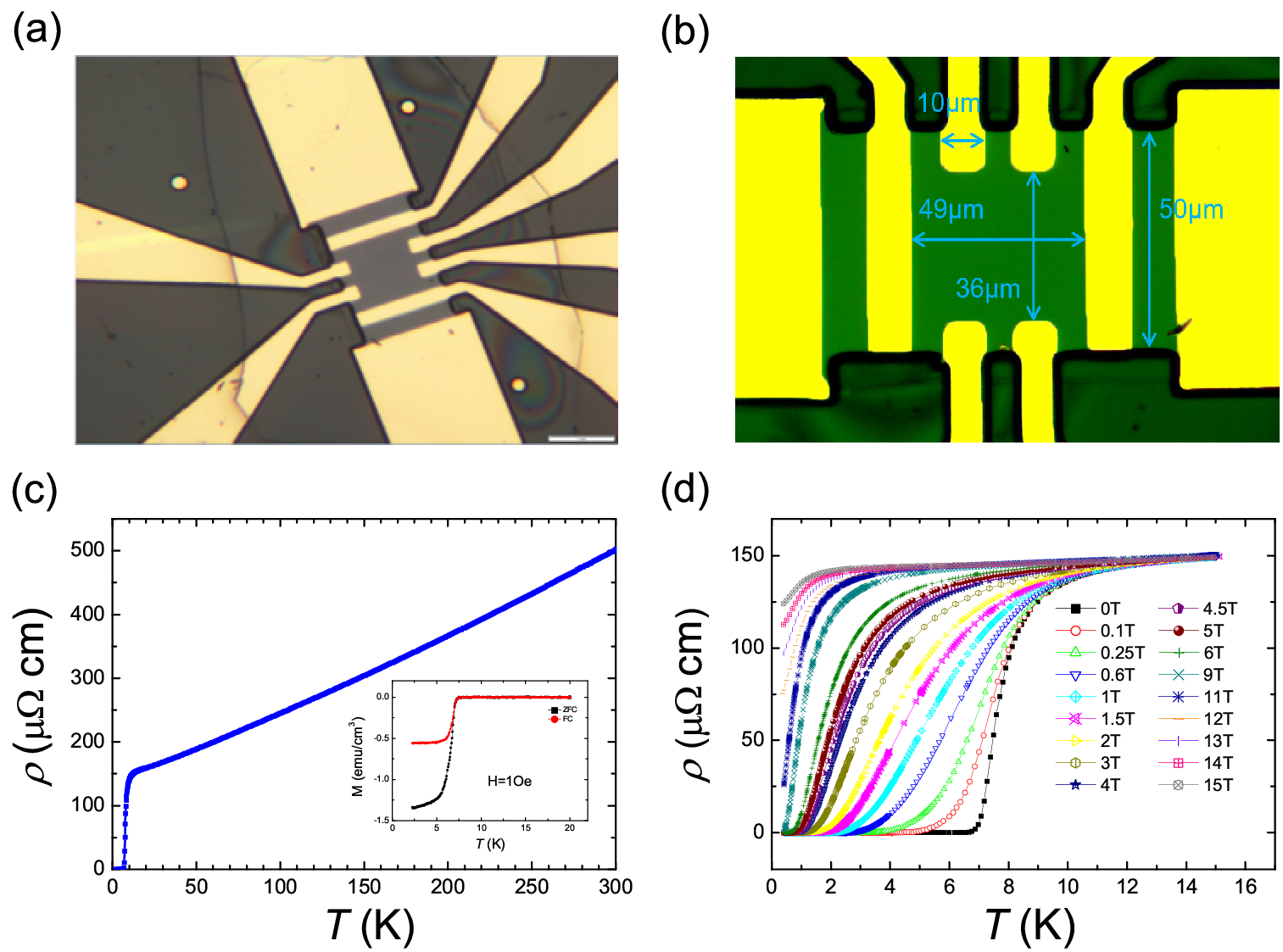}
\caption{ (a) Optical image of the Hall bar device made on our Bi2201 crystal. The yellow areas show the gold electrodes, and the black areas are the sapphire substrates exposed after the sample was etched away. The dark-grey bar in the middle represents the sample. (b) An enlarged optical image of the micro-bridge of Bi2201, the sizes of the sample and electrodes are also shown. Due to the different imaging technique used, now the sample is shown by dark-green color in the middle. (c) Temperature dependence of resistivity in a wide temperature region under zero applied field. The inset in Fig.~\ref{fig1}(c) shows the temperature dependence of magnetization measured under 1 Oe with the zero-field-cooling and field-cooling mode. (d) Temperature dependence of resistivity in the low temperature region under different magnetic fields.}
\label{fig1}
\end{figure}

Fig.~\ref{fig1}(a) and (b) show optical images of our Bi2201 sample with a Hall bar structure after the micro-fabrication. In Fig.~\ref{fig1}(b), the dark-green area denotes the Bi2201 sample, and the yellow areas represent the gold electrodes. The sizes of the sample and gold electrodes are also denoted in Fig.~\ref{fig1}(b). The thickness of the Bi2201 sample is $d = 1.5 $ $\mu$m measured by an atomic force microscope (AFM). We mark all the sizes here for determining the longitudinal resistivity and Hall coefficient. Since the electrodes have certain sizes, we use the central point to calculate related quantities. The well measured sizes allow us to precisely determine the absolute values of the resistivity and Hall coefficient.

Fig.~\ref{fig1}(c) shows the temperature dependence of the in-plane resistivity of Bi2201 at zero field in a wide temperature region, and zero resistance state is achieved at $T_{c}^{zero}$ = 6.7 K. In Fig.~\ref{fig1}(d), we present the resistive transitions below 15 K under different magnetic fields.  According to previous studies, with the nominal composition x = 0.05 and the $T_c$ value, we judge that the hole doping level of the present sample is in the region $0.14<p<0.16$, namely it is in the slightly underdoped region but close to the optimal doping point~\cite{LuoHQ}. The transition near the onset temperature is very rounded, indicating a strong superconducting fluctuation. This rounded resistive transition from temperatures above $T_c$ is a quite common effect in hole-doped cuprates, and is proved to arise from strong superconducting fluctuations by the observation of, for example, strong Nernst effect in La$_{2-x}$Sr$_x$CuO$_{4-\delta}$ and La-doped Bi2201~\cite{WangYYPRB}, and strong excess conductivity in YBa$_2$Cu$_3$O$_{7-\delta}$~\cite{Leridon}. Furthermore, this seemingly broad transition occurs mainly due to the gradual dropping down of resistivity in the temperature region far above $T_c^{zero}$, which occurs also in the original single crystals~\cite{LuoHQ}. This is not due to inhomogeneity as occurring in conventional superconductors. The basic reason is that the transition near zero resistance is sharp, which would not be the case if the sample was inhomogeneous. Thus it is hard to define an onset transition temperature $T_c^{onset}$ and determine the transition width. The inset in Fig.~\ref{fig1}(c) shows the temperature dependence of magnetization measured on one sample taken from the same batch under a field of 1 Oe. The onset transition of magnetization is about 7.1 K, corresponding roughly to the middle transition point of resistivity at zero field. One can see that, the resistivity exhibits a roughly $T$-linear dependence from 300 K down to 20 K. Below 20 K, there is a slight upward curvature of resistivity, but still a linear-like temperature dependence can be seen even down to 2 K after applying a magnetic field of 15 T. This slight upward curvature is induced by the elastic scattering of impurities, which will be addressed below. It is found that a magnetic field of 15 T can already kill the zero resistance state and recovers about 88\% normal state resistivity at 400 mK. This allows us to explore the most $H-T$ phase diagram.

\begin{figure}[H]
\centering
\includegraphics[width=7cm]{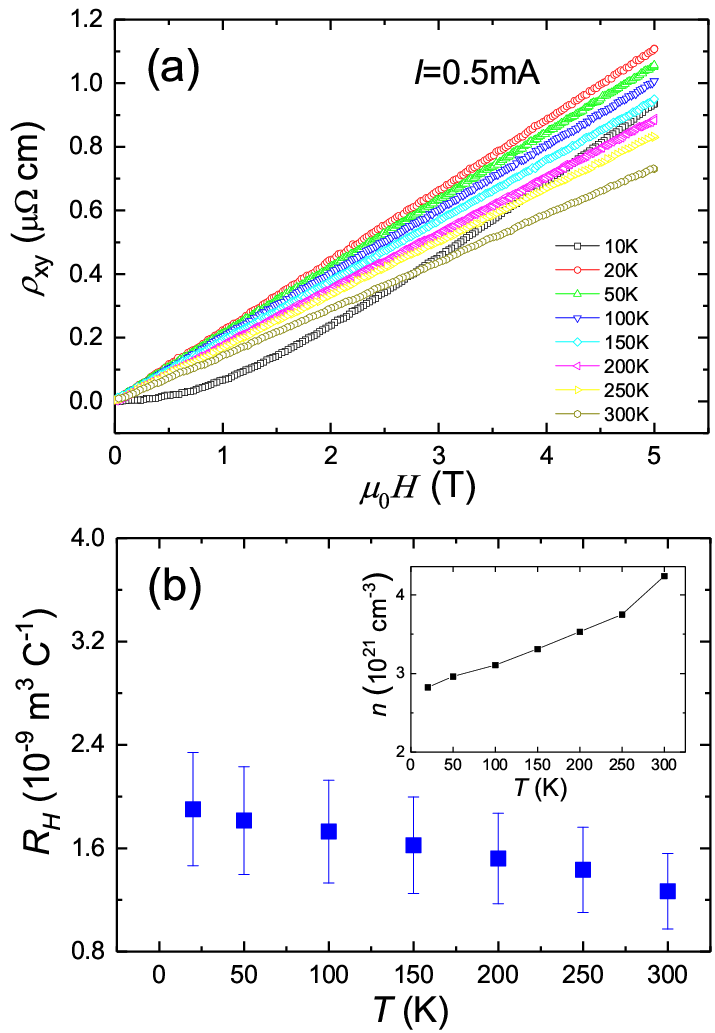}
\caption{ (a) Hall resistivity of the Bi2201 sample at different temperatures. Above 20 K, the Hall resistivity exhibits a linear relation with applied magnetic field, and the slope gives the Hall coefficient. (b) Hall coefficient $R_H$ of the Bi2201 sample as a function of temperature.  Inset: The carrier density $n$ obtained from the Hall coefficient $R_{H}$. Above 20 K, the carrier density $n$ changes mildly with temperature. The error bar is given here because the electrodes for the Hall voltage measurement have certain sizes.}
\label{fig2}
\end{figure}

Although the normal states of some cuprate superconductors exhibit anomalous behaviors, such as the $T$-linear dependence of resistivity and pseudogap effect, the quasi-particle features were still observed both in the normal state~\cite{Sato,TailleferPRB2021} and superconducting state~\cite{Hoffman,GuQQSTM}. Thus in the following, we use the Drude formula to extract the scattering rate, as described in reference~\cite{Planckian}. As adopted by many researchers, we assume the Drude formula can still be used to describe the quasiparticle scattering and transport properties in normal state, thus the resistivity can be written as,
\begin{equation}
\rho=\frac{m^{*}}{ne^{2}}\frac{1}{\tau}.
\label{equ1}
\end{equation}
Here $e$ is the charge of an electron, $n$ is the carrier density and $m^{*}$ is the effective mass of the quasiparticles, and the latter two parameters should be determined from experiment. The temperature dependence of resistivity is then reflected by the scatting rate $1/\tau$. In principle, the value of $m^{*}/n$ can be determined directly from the optical conductivity~\cite{LCCOGreene}, while due to the small Drude weight and very strong phonon contributions in the sample, we didn't successfully get credible data of $m^{*}/n$. Alternatively, we can get the charge carrier density from the Hall effect measurements. Fig.~\ref{fig2}(a) shows the Hall resistivity as a function of magnetic field at different temperatures. It is found that, above 20 K, the Hall resistivity changes linearly with magnetic field. At 10 K, the Hall resistivity shows a non-linear dependence of magnetic field, this may be induced by the involvement of vortex motion. We can get the Hall coefficient $R_{H}$ by linearly fitting to the Hall resistivity data in Fig.~\ref{fig2}(a). The obtained Hall coefficient $R_H$ is plotted as a function of temperature in Fig.~\ref{fig2}(b). One can see that there is a slight change of Hall coefficient $R_H$ above 20 K. The charge carrier density $n$ is obtained from the Hall coefficient $R_{H}$ based on the formula $R_{H} = 1/ne $. Since the temperature dependence of the carrier density $n$ is weak, we take the average of the carrier densities between 20 K and 300 K and get $n_{A} = 3.4\times10^{21}$ cm$^{-3}$. This value is close to the previously reported one in La-doped Bi2201~\cite{Bi2212}.

\subsection{Effective mass}\label{sec:3.2}

\begin{figure}[H]
\centering
\includegraphics[width=7cm]{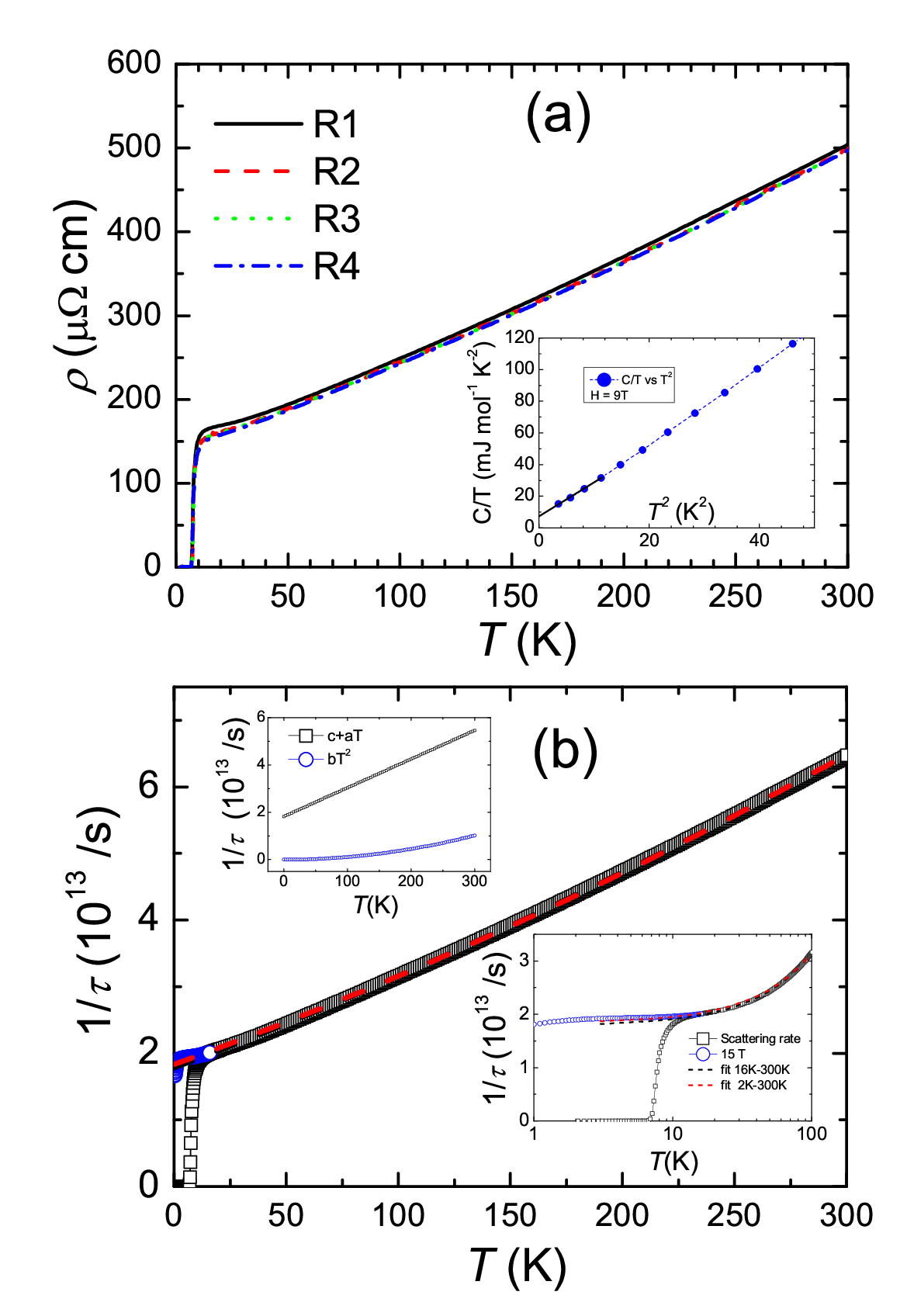}
\caption{ (a) Resistivity as a function of temperature for the Bi2201 sample measured at different time. Here R1, R2, R3 and R4 represent the measurements in different rounds. (b) Temperature dependence of scattering rate calculated by using the data of R4 and that under 15 T. The dashed lines are fits of the scattering rate from 16 K up to 300 K (black), 2 K up to 300 K (red) using the formula $1/\tau=aT+bT^{2}+c$. The scattering rate is obtained from resistivity data, via Eq.~\ref{equ1}. The right-bottom inset of (b) shows an enlarged view of the scattering rate from 1 K up to 100 K in a semi-logarithmic scale. The upper-left inset of (b) shows a separation of the resistivity into a linear term and quadratic term according to the fitting.}
\label{fig3}
\end{figure}

Now let us turn to the effective mass of Bi2201. Generally speaking, it is quite hard to obtain the effective mass of a correlated metal. Usually, one can estimate the effective mass from measurements of quantum oscillation, optical reflectivity or specific heat. Here we use specific heat to determine the effective mass $m^{*}$ by using the relation~\cite{effective-mass} $\gamma_n = (\pi N_{A}k_{B}^{2}/3\hbar^{2})a^{2}m^{*}$, where $a$ is the lattice constant, $N_{A}$ is the Avogadro's number. In the paper of Legros \textit{et al.}~\cite{Bi2212}, concerning the effective mass of Bi2201, they referred to the La-doped Bi-2201 with $T_c$ = 19 K in the overdoped regime~\cite{specific-heat}. About the La-free Bi2201, there have been some published results of specific heat, but mainly concerning the low lying quasiparticle excitations in superconducting state~\cite{Bi2201SH1,Bi2201SH2}. Thus there have been no reported values about $\gamma_n$ until a recent report~\cite{Girod}. In this paper, except for a collection and analysis of specific heat for La-doped Bi2201 and La$_{2-x}$Sr$_x$CuO$_4$, the authors present one set of data for the La-free Bi2201 sample with doping level close to ours (they called Bi2201 \#1). They found that the $C/T$ exhibits a saturation and slight upturning in double logarithmic plot of $C/T$ vs. $T$ in low-$T$ region. This slight upturning makes it difficult to precisely determine the $\gamma_n$ value, a gross estimate of $\gamma_n$ at 0.65 K is about 13 $\pm$ 1 mJ mol$^{-1}$K$^{-2}$. For comparison, we took a La-free Bi2201 sample with similar doping level and $T_c$ as the one for our transport measurements to measure the specific heat at low temperatures. The measured data at 9 T down to 1.89 K are shown in the inset of Fig.~\ref{fig3}(a). It is found that the data can be roughly fitted with the Debye model in low-$T$ region. If we use a linear extrapolation of the low-$T$ data in the form of $C/T$ vs. $T^2$, as highlighted by the solid line in low-$T$ region in the inset of Fig.~\ref{fig3}(a), we get an extrapolated value of $\gamma$ (9T) = 7.4 mJ mol$^{-1}$K$^{-2}$ at $T = 0$ K. The values of $C/T$ of our results are grossly consistent with the data of Girod \textit{et al.}~\cite{Girod} at temperatures above 2 K. Although the upper critical field at $T = 0$ K is higher than 9 T in the present system, from the resistivity data, one can see that a field of 9 T has already killed the zero resistance state above 0.4 K, thus even $\gamma$ will still increase with field beyond 9 T, but that increase should be limited. Thus, for the La-free Bi2201 near the optimal doping point with $T_c$ of about 6.7-10 K (note: different people define $T_c$ with different criterions, with doping level in the region 0.13 $< p <$ 0.16), the $\gamma_n$ may locate in the region of 8-13, or 10.5 $\pm$ 2.5 mJ mol$^{-1}$K$^{-2}$ (the upper bound was quoted from Fig.6 of Girod \textit{et al.}~\cite{Girod}). By taking this value we can estimate $m^{*} = 7.35\pm1.75 m_{0}$ via the equation between $\gamma_n$ and $m^*$, where $m_{0}$ is the bare electron mass.

Fig.~\ref{fig3}(a) displays the resistivity data measured at different time. The measurements of Round 2 (abbreviated as R2) and R3 were carried out half a year after the first measurement R1. The measurement of R4 was carried out nine months after the first measurement. There is only a slight change of resistivity in the low temperature region with measurements spanned in about nine months, but the transition part overlaps very well, so the Bi2201 sample is quite stable. The resistivity plotted in Fig.~\ref{fig1}(c) is R4 displayed in Fig.~\ref{fig3}(a). Using $n_{A}=3.4\times10^{21}$ cm$^{-3}$ determined from the Hall effect measurements, and $m^{*} = 7.35 m_{0}$, we calculate the scatting rate as a function of temperature for our Bi2201 sample from the R4 displayed in Fig.~\ref{fig3}(a) via Eq.~\ref{equ1}. Shown in Fig.~\ref{fig3}(b) are the scattering rates derived from the data R4 (open dark symbols) together with that measured under 15 T (blue circles) at low temperatures. Perhaps due to the slightly different electric connection details in different equipments, there is a little difference between these two sets of data, thus the data measured under 15 T was off-set up of about 3\% in order to have a smooth connection at 16 K and above. We should note that the residual resistance ratio RRR = $\rho(300K)/\rho(0K)$ $\approx$ 3.33 in present sample is not big. This could be induced by the larger out-of-plane disorder effect induced by a larger mismatch of the ionic radius between Sr$^{2+}$ and Bi$^{3+}$. Since the magnetoresistance at higher temperatures, for example at 16 K, is very small, we believe the 3\% mismatch of resistivity under 15 T (measured in Pekin University by another PPMS) is not due to the disorder scattering. And the temperature dependent scattering rate is intrinsic and reflects the electronic properties of the system.

\begin{table}[H]
\centering
\caption{Results of fitting parameters for the scattering rate in different temperature ranges}\label{tab1}
\doublerulesep 0.1pt \tabcolsep 11pt
\begin{tabular}{cccc}
\hline
   Range  & $a(*10^{10})$ & $b(*10^{10})$  & $c(*10^{10})$  \\
\hline
16K-300K &   $12.78$ &  $0.951*10^{-2}$   &   $1782$ \\
\hline
2K-300K &   $12.11$ &   $1.135*10^{-2}$  &   $1832$  \\
\hline

\end{tabular}
\end{table}

The dashed lines plotted in Fig.~\ref{fig3}(b) are the fitting curves of scattering rate in different temperature regions using the formula $1/\tau=aT+bT^{2}+c$, where $a$, $b$ and $c$ are the fitting parameters. In order to see the fitting more clearly at low temperatures, we enlarged the view by taking a semi-logarithmic scale (right-bottom inset of Fig.~\ref{fig3}(b)). One can see a little difference between the data and the fitting curves at low temperatures below about 5 K, which may be induced by the presence of small amount of elastic scattering. The global fittings look quite good and the fitting parameters are given in Table I. In the upper-left inset of Fig.~\ref{fig3}(b), we show the contributions of the $T$-linear and quadratic temperature dependence arising from the fitting to the data between 2-300 K, and one can see that the quadratic term is much smaller than the linear term, especially below 200 K. It is clear that the fitting parameters are quite close to each other when we fit the data in temperature regions between 16-300 K or 2-300 K. For temperatures ranging from 2 K to 300 K, the linear term can give us $\alpha = 0.925$ from the fit. Taking the different Hall coefficient values at 20 K and 300 K into consideration, we find that $0.77<\alpha<1.16$. Therefore, our refined fitting results confirm the Planckian dissipation over a wide temperature range~\cite{Davison,Hartnoll,SKY}. We need to notice that, in recent publications~\cite{Barisic,Greven}, specially in the system HgBa$_2$CuO$_{4+\delta}$, it was shown that either the longitudinal resistivity $\rho(T)$ or the Hall angle $cot(\Theta_H)=\rho(T)/\rho_H\propto m^{*}/\tau$ exhibits a quadratic temperature dependence in low-$T$ region, suggesting a Fermi liquid feature. We tried to plot our data either in $\rho(T)$ or $cot(\Theta_H)$ versus $T^2$, but we didn't see a quadratic temperature dependence in low-$T$ region. Although the curve $cot(\Theta_H)\propto m^{*}/\tau$ versus $T$ shows a little curvature in the high temperature region, the data in low-$T$ region (up to about 65 K) can still be fitted with a linear relation. Thus we believe the scattering rate of present Bi2201 sample does not have a quadratic temperature dependence in low-$T$ region, which is different from the HgBa$_2$CuO$_{4+\delta}$.

\subsection{Upper critical field}\label{sec:3.3}

\begin{figure}[H]
\includegraphics[width=8cm]{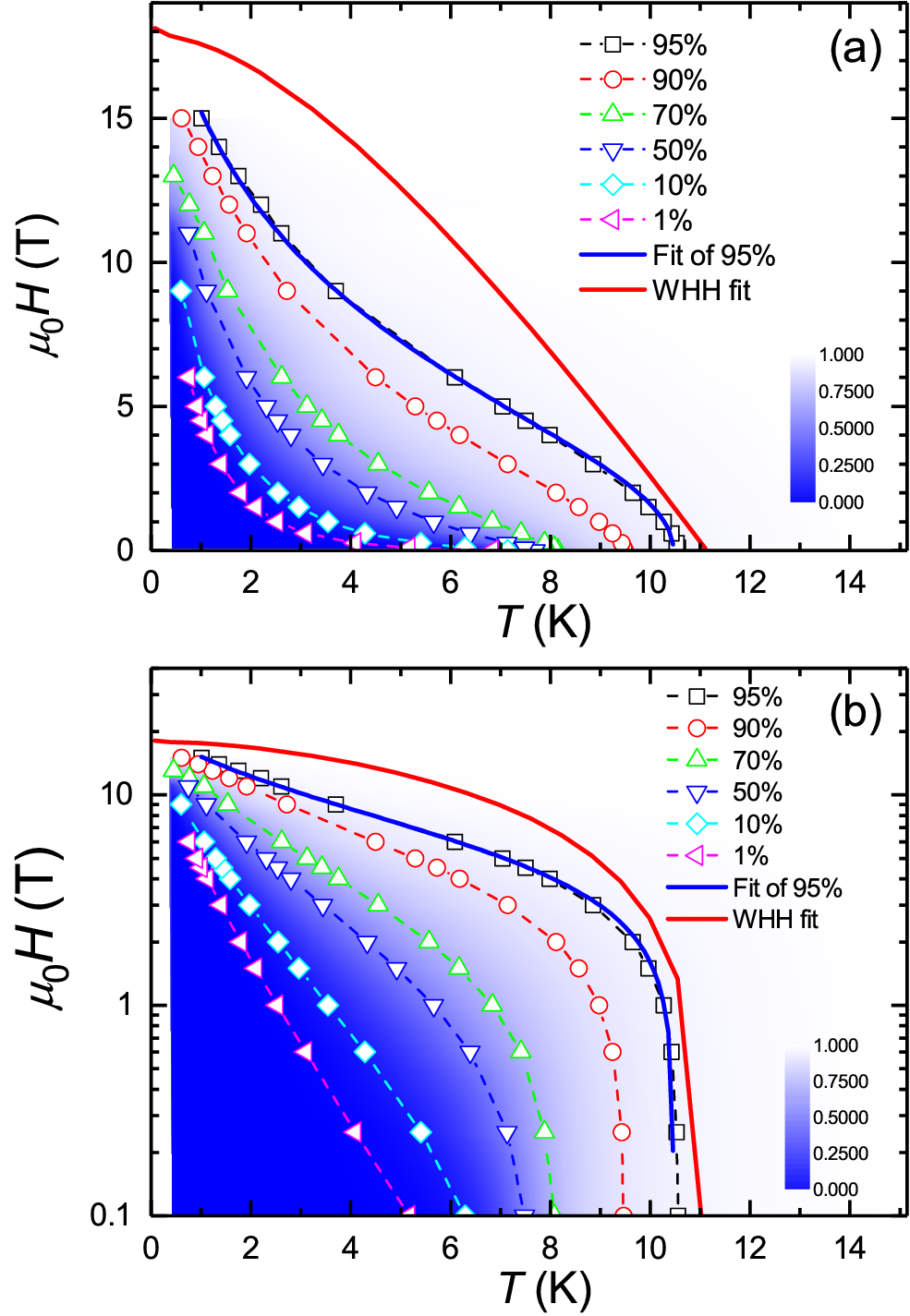}
\caption{ (a) Phase diagram of the upper critical field $H_{c2}(T)$ obtained from Fig.~\ref{fig1}(d) by using different criterions. The color intensity reflects the ratio of $\rho(T,H)/\rho_n(T)$, as marked by the color bar. The $H_{c2}(T)$ values are obtained from the magnetic fields at which the resistivity has reached $1\%$, $10\%$, $50\%$, $70\%$, $90\%$ and $95\%$ of its normal-state values. The dashed lines are guides to the eyes. The red solid line shows the fitting result based on the WHH theory. The blue solid line displays the fitting result according to the model (Eq.~\ref{equ3}). (b) The phase diagram $H_{c2}(T)$ with semi-logarithmic scale. The physical meaning of the data and the lines are the same as in (a). }
\label{fig4}
\end{figure}

Fig.~\ref{fig4}(a) shows the temperature dependencies of the critical field at which the resistivity has reached 1$\%$, 10$\%$, 50$\%$, 70$\%$, 90$\%$ and 95$\%$ of $\rho_n(T)$. Here the normal state resistivity $\rho_n(T)$ is obtained by following the normal state extrapolation linear line measured between 10 and 20 K under a high magnetic field. The same data are shown in Fig.~\ref{fig4}(b) in a semi-logarithmic way. We choose the field at which the resistivity has reached 95$\% \rho_n$ as the onset point of upper critical field $H_{c2}(T)$, thus it is called as $H_{c2}^{on}(T)$. In the following we focus on fitting the $H_{c2}^{on}(T)$ curve according to the formula expected by the Ginzburg-Landau (GL) theory, since this line is the usually defined upper critical field at which the droplet of superconductivity starts to form. One can see that the $H_{c2}^{on}(T)$ displayed in Fig.~\ref{fig4}(a) does not saturate in the zero temperature approach, which violates the expectation by the GL theory. The critical fields determined with other criterions of lower resistivity ratio $\rho(T)/\rho_n(T)$ should have involved the flux flow, thus its divergent behavior at low temperatures is understandable. In order to demonstrate that the $H_{c2}^{on}(T)$ curve strongly deviates from the behavior expected by the GL theory, we calculated the $H_{c2}(T)$ curve based on the Werthamer-Helfand-Hohenberg (WHH) theory~\cite{WHH}. In the dirty limit which is the case for the Bi2201 system, the upper critical field can be described by the WHH theory via the function~\cite{WHH}
\begin{equation}
\begin{split}
\ln \frac{1}{t}=\sum_{\nu=-\infty}^\infty \Big\{ \frac{1}{|2\nu + 1|} - \Big[{|2\nu + 1|} +\\
     \frac{\bar{h}}{t} + \frac{(\alpha_0\bar{h} / t)^{2}}{|2\nu + 1|+(\bar{h}+\lambda_{so}) / t}\Big]^{-1}\Big\}.
\end{split}
\label{equ2}
\end{equation}
Here, $t = T/T_{c}$, $\bar{h} = 4\mu_{0}H_{c2}/(\pi^2H'T_c)$ with $H' = \mu_{0}|dH_{c2}/dT|_{T_{c}}$; $\lambda_{so}$ is the parameter representing the strength of the spin-orbit interaction; $\alpha_0$ is the one reflecting the strength of the spin paramagnetic effect. The red solid lines displayed in Fig.~\ref{fig4}(a) and (b) are the fitting results with the parameters $\alpha_0=0$ and $\lambda_{so}=0$ according to the WHH theory. The fitting result has no way to be consistent with the experimental results. The fitting quality is equally poor when we adjust the parameters $\alpha_0$ and $\lambda_{so}$ (not shown here). The difficulty for the fitting here is that all the theoretical curves should show a negative curvature and flattened feature in the low temperature limit, but the experimental data all show a positive curvature and a clear divergent behavior. We should note that, in the cuprates, this kind of upward curvature of $H_{c2}(T)$ in low-$T$ region has been observed in many other systems~\cite{TlBaCuO,AndoPRB,Osofsky,Sebastian,WenHHPRL1999}. It was argued that this kind of divergence of $H_{c2}(T)$ may have involved the vortex motion~\cite{WenHHPRL1999,WenHHPRL2000}.

\section{Discussion}\label{sec:4}

We have shown the dominant $T$-linear resistivity in the ``normal state'' and the anomalous behavior of the upper critical field $H_{c2}$ defined in the usual way in La-free Bi2201. The ``normal state'' is abnormal with a dominant $T$-linear resistivity, this occurs when the external field reaches the so-called $H_{c2}$. Thus these two effects should be naturally connected to each other. Since the resistivity data shows an asymptotical approach to the normal state value when temperature is increased under a certain magnetic field, it seems quite hard to precisely define at which point the resistivity starts to deviate from the normal state value $\rho_n(T)$. This has been addressed previously~\cite{GrissonnancheNC,RamshawPRB} and it seems always debatable in cuprate superconductors. In the study of Grissonnanche \textit{et al.}~\cite{GrissonnancheNC}, the authors comparatively measured thermal conductivity and resistivity versus magnetic field at a certain temperature, and found that the upper critical field determined by using a criterion of high ratio of $\rho(T)/\rho_n(T)$ coincides with the field at which a sharp drop of thermal conductivity was observed. This suggests that the upper critical field determined here may reflect the ending point of the Abrikosov vortex state. However, it remains to know whether it also corresponds to the pair breaking field. Thus we also use a high resistivity ratio $\rho(T)/\rho_n(T)$, here for example 95\% $\rho_n(T)$, to define the upper critical field. The obtained $H_{c2}(T)$ data show an upward curvature in the low temperature limit. Actually, Vedeneev \textit{et al.}~\cite{Hc2normal} have measured the upper critical field in the Bi2201 system by using a criterion with even a higher ratio of resistivity, they still found this type of divergence of $H_{c2}(T)$ in low-$T$ region. This already hinges on that the upper critical field determined by the usually adopted way does not reflect the real pair breaking field. The strong Nernst signal detected far above the usually defined $H_{c2}(T)$ can corroborate this picture~\cite{XuZANature,WangYYPRB}. It is clear that the superconducting fluctuation should be very strong in the present sample as evidenced by the very rounded transition near the onset transition temperature. For example, the resistivity shown in Fig.~\ref{fig1}(d) starts to drop from the high temperature background at about 15 K, but the zero resistance is achieved at about 6.7 K. As said before, this broad transition is not due to inhomogeneous feature as that in conventional superconductors, since the foot of the resistive transition at zero field is very sharp. In this case, if we take a criterion 95$\%$ of $\rho_n(T)$ to define the so-called upper critical field, we should have encountered a state with many uncondensed Cooper pairs. It was proposed by Alexandrov \textit{et al.} that~\cite{Alexandrov1}, for a critical field marking the resistive transition of uncondensed charge bosons, the critical field should hold a power-law function of $1-t$ in a wide temperature region due to the variation of the superfluid phase stiffness, with $t=T/T_c^{on}$ and $T_c^{on}$ the onset transition temperature; and an inverse power-law function of temperature in the low temperature region. Thus we propose a general formula for the ``upper critical field'' based on this idea, which reads as
\begin{equation}
H_{c2}(T)=H^*(1-t)^p/(t+\delta)^q,
\label{equ3}
\end{equation}
where $H^*$ and $T_{c}$ are parameters which can be roughly predetermined by extrapolations; $p$, $q$ and $\delta$ are fitting parameters, and $\delta$ gives a slight modification to the denominator $t^q$ and should be much less than 1. We thus fit our experimental data to this equation. The blue solid line shown in Fig.~\ref{fig4}(a) and (b) denotes the fit of $H_{c2}^{on}(T)$ based on our proposed formula, the best fit yields $p$ = 0.5, $q$ = 0.5, $\delta$ = 0.154 with $H^*$ = 8 T and $T_{c}$ = 10.46 K. It seems that the model can fit the data rather well. A nice fit to the data with 50$\%$ of $\rho_n(T)$ was also achieved with $p$ = 0.88, $q$ = 1.38, $\delta$ = 0.305 with $H^*$ = 3.43 T and $T_{c}$ = 7.7 K. Since now flux flow is heavily involved in the dissipation and more uncondensed Cooper pairs (less quasiparticles) are concerned, it is natural to see different fitting parameters. We also use the model to fit the resistive upper critical field of Tl$_{2}$Ba$_{2}$CuO$_{6}$ reported by Mackenzie $et$ $al$.~\cite{TlBaCuO} and the fitting result is also perfect, yielding the values of $p$ = 2, $q$ = 0.5, $\delta$ = 0.024 with $H^*$ = 2.39 T and $T_{c}$ = 16.8 K. Thus we believe the model, although crude in its form, captures the major physics of the ``upper critical field'' of the cuprate system. Recalling the original meaning with this equation, we argue that the ``normal state'' reflects a mixture of uncondensed Cooper pairs and strongly renormalized quasiparticles. This conclusion is close to the picture of pair density wave emerging in the vortex halo of cuprate superconductors~\cite{PDWSTM,PDWSummary}.

\section{Conclusion}\label{sec:5}

In conclusion, we have carried out resistivity and Hall effect measurements of Bi2201 single crystal on a micro-fabricated bridge under magnetic fields. Application of a 15 T magnetic field can successfully suppress the superconductivity and recover about 88\% ``normal state'' resistivity at 400 mK. The resistivity above about 2 K shows a dominant linear temperature dependence, mixed with a residual term in low-$T$ region and a small quadratic term. From this dominant $T$-linear term, together with the Hall coefficient and effective mass estimated from specific heat, we determined the scattering rate and confirmed the scenario of Planckian dissipation. The determined upper critical field $H_{c2}(T)$ by using a criterion of 95\% $\rho_n$ exhibits an anomalous positive curvature in the low temperature approach. Based on the picture of charged bosons formed above $T_c$, we proposed a formula for describing the data of $H_{c2}(T)$. Our results strongly indicate that the $T$-linear resistivity in the ``normal state'' and the anomalous $H_{c2}(T)$ curve may be tightly linked each other, which suggests a ``normal state'' with uncondensed Cooper pairs and heavily renormalized quasiparticles.

\begin{acknowledgments}
We acknowledge Jan Zaanen, Richard Greene and Ilya Eremin for the useful discussions. This work was supported by the National Natural Science Foundation of China (Grant Nos: 11927809, NSFC-DFG12061131001, 61727805, 11888101) and National Key R and D Program of China (Grant Nos. 2016YFA0300401, 2021YFA0718802, 2018YFA0305604), and the Strategic Priority Research Program of Chinese Academy of Sciences (Grant No.XDB25000000), the Beijing Natural Science Foundation (Grant No. Z180010).
\end{acknowledgments}





\end{document}